%% file: expertise.tex
\documentclass[sigconf]{acmart}
\usepackage{xcolor}
\usepackage{hyperref}

\usepackage{balance}
\definecolor{main}{HTML}{5989cf}    % setting main color to be used
\definecolor{sub}{HTML}{cde4ff}     % setting sub color to be used

\definecolor{firebrick}{HTML}{B22222}
\definecolor{yellowgreen}{HTML}{9ACD32}
\definecolor{royalblue}{HTML}{4169E1}

\begin{document}
\title[U Owns the Code That Changes]{U Owns the Code That Changes and How Marginal Owners Resolve Issues Slower in Low-Quality Source Code}
%\title[Green Devs in Red Code]{Green Devs in Red Code: U Owns the Code and its Impact on Issue Resolution Time in Low-Quality Source Code}
%\subtitle{}

\author{Markus Borg}
\orcid{XXX}
\affiliation{%
  \institution{CodeScene AB and Lund University}
  \city{Malmö}
  \country{Sweden}
}
\email{markus.borg@codescene.com}

\author{Adam Tornhill}
\orcid{XXX}
\affiliation{%
  \institution{CodeScene AB}
  \city{Malmö}
  \country{Sweden}
}
\email{adam.tornhill@codescene.com}

\author{Enys Mones}
\orcid{XXX}
\affiliation{%
  \institution{CodeScene AB}
  \city{Malmö}
  \country{Sweden}
}
\email{enys.mones@codescene.com}

\renewcommand{\shortauthors}{Borg \textit{et al.}}

\begin{abstract}
[Context] Accurate time estimation is a critical aspect of predictable software engineering. Previous work shows that low source code quality increases the uncertainty in issue resolution times.
[Objective] Our goal is to evaluate how developers' project experience and file ownership are related to issue resolution times.
[Method] We mine 40 proprietary software repositories and conduct an observational study. Using CodeScene, we measure source code quality and active development time connected to Jira issues. 
[Results] Most source code changes are made by either a marginal or dominant code owner. Also, most changes to low-quality source code are made by developers with low levels of ownership. In low-quality
source code, marginal owners need 45\% more time for small changes, and 93\% more time for large changes.
[Conclusions] Collective code ownership is a popular target, but industry practice results in many dominant and marginal owners. Marginal owners are particularly hampered when working with low-quality source code, which leads to productivity losses. In codebases plagued by technical debt, newly onboarded developers will require more time to complete tasks.
\end{abstract}

\copyrightyear{2023} 
\acmYear{2023} 
\setcopyright{acmcopyright}
\acmConference[EASE 2023]{The 24th International Conference on Evaluation and Assessment in Software Engineering}{14–16 June, 2023}{Oulu, Finland}
\acmPrice{15.00}
\acmDOI{10.1145/3522664.3528592}
\acmISBN{978-1-4503-9275-4/22/05}

\keywords{mining software repositories, source code quality, code ownership, issue resolution time, technical debt}

\maketitle

\input{body.tex}

\balance
\bibliographystyle{ACM-Reference-Format}
\bibliography{expertise}

\end{document}

%% file: body.tex
\section{Introduction}
Predictability is a crucial factor in successful software engineering.
By allowing teams to plan and manage their work effectively, predictability helps organizations deliver high-quality products on time and within budget~\cite{boehm_seven_1983,naedele_manufacturing_2015,malgonde_ensemble-based_2019}.
This is accomplished through the ability to estimate the time and resources required for a project, identify potential challenges and roadblocks, and make informed decisions.
Furthermore, predictability enables organizations to identify areas for improvement and implement changes to increase efficiency and productivity. 
Accurate time estimation is a critical aspect of predictable software engineering.
Numerous researchers have proposed ways to provide estimates~\cite{jorgensen_systematic_2007}.
However, providing accurate estimates is notoriously difficult due to the inherent uncertainty of software development.

High source code quality decreases the uncertainty in issue resolution times.
Our previous work highlights the substantial variability in issue resolution times in low-quality code~\cite{tornhill_code_2022}.
In a large-scale study of 39 proprietary codebases, we found that the average maximum resolution time in source code files with severe maintainability problems was nine times higher compared to high-quality counterparts.
Based on this finding, we concluded that inadequate source code quality challenges issue resolution predictability.

In this paper, we investigate a possible reason for the variability.
We hypothesize that the variability can partly be explained by the \textit{project experience} and \textit{code ownership} of the developer resolving the issue.
Using the terminology from the Code Red paper~\cite{tornhill_code_2022}, perhaps green developers suffer more from red code?
We use CodeScene's Code Health metric to measure code quality and the tool's definition of time-in-development to measure issue resolution time.
In this study, we study how Code Health and time-in-development relate to project experience and code ownership.

%We conduct repository mining that matches the size of Tornhill and Borg's study.
Our dataset comprises 48,644 \textit{touches}\footnote{A touch is a modification to a single file~\cite{foucault_usefulness_2015}, see Fig.~\ref{fig:data_coll}.} to 13,433 source code files from 40 proprietary projects.
The data is augmented by 7,307 Jira issues used to calculate time-in-development.
We find that the ownership distribution of touched source code files has a clear U-shape, explaining the title of this paper.
Moreover, most changes to low-quality source code are made by developers with low ownership. Marginal owners also resolve issues slower in low-quality source code -- the median time for small and large changes is 45\% and 93\% longer, respectively.

Our study brings a novel perspective on proprietary software engineering at a rare scale.
Few repository mining studies target tens of closed-source projects and the distribution of programming languages contrasts most academic studies.
Based on our large-scale study, we make two main conclusions.
First, we find that the agile practice of collective code ownership has limited influence on proprietary codebases.
In line with previous work, we find a relationship between low ownership and poor source code quality.
Second, we provide additional evidence that low-code quality hampers development productivity.
The technical debt's interest is higher for developers working with less familiar source code.
Thus, technical debt impacts the onboarding of developers and introduces predictability risks for anyone working on previously unfamiliar code. This makes it harder for organizations to actively manage knowledge distribution within development teams.
%While development organizations certainly should strive to pay off the technical debt in evolving source code, they should also manage knowledge distribution accordingly.

The paper is organized as follows. Sec.~\ref{sec:bg} introduces the concept of collective code ownership and the two metrics Code Health and time-in-development.
Sec.~\ref{sec:rw} shares related work on source code quality versus ownership and experience.
Sec.~\ref{sec:method} describes how we conduct repository mining and analyze the data.
In Sec.~\ref{sec:res} we present our results and discuss their implications.
Sec.~\ref{sec:threats} elaborates on threats to validity before we conclude the paper in Sec.~\ref{sec:conc}.

\section{Background} \label{sec:bg}
This section introduces the agile practice of collective code ownership, and the two metrics Code Health and time-in-development. 

Collective code ownership is a key concept in agile software development~\cite{abrantes_common_2011}. The idea is to avoid the situation where every piece of source code has an ``owner'' responsible for its maintenance. According to the Agile Alliance's Agile Glossary, collective code ownership means that every developer ``has a positive duty to make changes to any code file as necessary.'' This concept is suggested to mitigate the risk of critical knowledge loss when developers quit, reduce bottlenecks caused by unavailable developers, neutralize Conway's law~\cite{bailey_decade_2013}, support knowledge transfer, and encourage developers to feel responsible for the quality of the entire system. Using grounded theory at Pivotal, Sedona \textit{et al.} reported that collective code ownership is a feeling that goes beyond a policy to be decreed~\cite{sedano_practice_2016}. When the feeling is present, Pivotal's developers found it highly advantageous, leading to team cohesion and improved system understanding -- and higher code quality.

Many researchers have proposed source code quality metrics~\cite{riaz_systematic_2009,baggen_standardized_2012}. In a 2017 systematic literature review, Nunez-Varela \textit{et al.} identified almost 300 metrics~\cite{nunez-varela_source_2017}. Object-oriented programming is by far the most studied paradigm in academic research and the most commonly used metrics are 1) the Chidamber and Kemerer metrics, 2) Lines of Code (LoC), 3) McCabe's cyclomatic complexity, and 4) the number of methods and attributes. In March 2021, the standard ISO/IEC 5055 Automated Source Code Quality Measures~\cite{international_organization_for_standardization_information_2021} was published as a way to evaluate the maintainability, reliability, security, and performance of systems already on the source code level, i.e., without executing the systems. The quality measures rely on counting occurrences of a controlled list of weaknesses, agreed upon as detrimental by the standardization initiative.

Code Health is a metric calculated by the CodeScene tool. The metric is a numeric and absolute score ranging from 10 (code of superior quality) to 1 (terribly bad code quality). Code Health acknowledges the idea that the best approach to measuring source code complexity is by identifying and measuring a set of specific complexity attributes~\cite{fenton_software_1994}. To meet this end, Code Health incorporates low-level elements from the ISO/IEC~5055 maintainability section, and complements them with design-level code smells\footnote{\url{https://codescene.io/docs/guides/technical/code-health.html}}. Examples of included design smells are God Class, God Methods, and Duplicated Code~\cite{lacerda_code_2020}.
Based on the Code Health scores, CodeScene categorizes source code files as either \textit{healthy} ($>8$), \textit{warning} (4--8), or \textit{alert} ($<4$). %CodeScene can be freely used for Open-Source Software (OSS) and research projects~\cite{tornhill_prioritize_2018}, and we rely on the tool for all software quality measurements in this study.

In this study, we are interested in development cycle times rather than lead times. In Jira terminology, cycle time is the time from the beginning to the end of a certain action, i.e., the time during which the issue has the status ``In progress''. Second, lead time covers the entire time from receiving a request for an action to the moment this action is completed, i.e., including the time in the queue. In this study, we represent cycle times using \textit{time-in-development} as defined in our previous work~\cite{tornhill_code_2022}, relying on a combination of transition timestamps in Jira and commit metadata from git. Cycle times differ from lead times by excluding the time involved in company-specific post-development steps, e.g., testing, release management, and deployment. As such, time-in-development eliminates several confounding factors that could influence the results of this study. What we measure as time-in-development is referred to as ``In-Progress Time'' in a recent study by Tawosi \textit{et al.}~\cite{tawosi_relationship_2022}.

\section{Related Work} \label{sec:rw}

%\subsection{Cost estimation}
%Grimstad and Jorgensen list seven factors that might lead to higher cost estimation errors~\cite{grimstad_framework_2006}: 1) project management ability, 2) developer skills, 3) performance of clients and sub-contractors, 4) completeness and certainty of information, 5) project complexity, 6) project priorities, and 7) project execution flexibility. The authors highlight that these factors reflect the ``cost factors'' in early work on software engineering economics by Boehm~\cite{boehm_software_1984}. Our current study is related to the second factor, i.e., developer skills.

%\subsection{Collective Code Ownership in Practise}
Several researchers have conducted surveys to explore how common different agile practices are in industry. In a survey of 109 experienced agile developers, Kurapati \textit{et al.} report that roughly 50\% of the respondents used collective code ownership~\cite{kurapati_agile_2012}. In a similar survey of 79 agile developers in Pakistan, Rauf and AlGhagees found that 43\% of the respondents use the practice~\cite{rauf_gap_2015}. On the contrary, in a larger study with 408 respondents in Finland, Rodriguez \textit{et al.} found that collective code ownership was the least used of the 16 studied agile practices~\cite{rodriguez_survey_2012}. The survey research shows that collective code ownership is applied in the software industry, but it is not one of the most common agile practices.

Sindhgatta \textit{et al.} conducted a case study at IBM investigating the evolution of an enterprise system adhering to agile methods~\cite{sindhgatta_software_2010}. A project team of 60 members was studied during 25 sprints over 15 months. Collective code ownership was studied as one out of several aspects. The authors found that ownership was successfully shared and 60\% of the files were owned by more than one developer.

%\subsection{Experience, Ownership, and Quality}    
Bird \textit{et al.} investigated the relationship between code ownership and software quality at Microsoft~\cite{bird_dont_2011}. Analyzing the development history of Windows Vista and Windows 7, they found that the number of minor contributors to a Windows binary strongly correlates with defects, even when controlling for size, churn, and complexity. Assuming that minor contributors have less expertise related to the corresponding files, the authors recommend more careful code reviews and discussing proposed changes with developers with a higher degree of ownership. 

Bird \textit{et al.}'s study has been replicated twice. First, Foucault \textit{et al.} replicated it for OSS development represented by four Apache projects and three Eclipse projects~\cite{foucault_code_2014}. However, despite using blocking to eliminate the impact of size, the authors did not find a correlation between file-level ownership and defects. Instead, they report that source code metrics are better than ownership for defect prediction. Second, Greiler \textit{et al.} conducted another replication, within Microsoft, investigating Office, Windows, Office365, and Exchange~\cite{greiler_code_2015}. Compared to Bird \textit{et al.}, the analysis was done on file- and directory level. The results confirm the original study, i.e., weakly owned files and directories correlate with defect proneness.

The correlation between ownership and defect proneness has inspired work on defect prediction. Jureczko and Madeyski studied 35 systems (5 industrial, 13 OSS, and 17 academic) and found ownership to be a relevant factor in improving prediction accuracy~\cite{jureczko_cross-project_2015}. Based on a study of nine OSS systems, Rajapaksha \textit{et al.} found that ownership above 85\% decreases the risk of a source code file being defective~\cite{rajapaksha_sqaplanner_2022}. Hattori \textit{et al.} challenges the predictive power of ownership by pointing at the imperfect memory of developers~\cite{hattori_refining_2012}: a developer who changed a file many times way back may be considerably less familiar with the source code compared to another developer who made few but recent changes. The authors mitigated the forgetting effect by refining the ownership metric by adding the time factor. 

Rahman and Devanbu studied the impact of code ownership and experience on software quality~\cite{rahman_ownership_2011}. They did not measure experience as a duration of time. Instead, they counted the total number of deltas committed by the author up to a particular point in time. Based on an analysis of four medium to large OSS projects, they found no association between experience and the number of bug resolutions addressing severe defects. In contrast to most previous work on ownership, we do not study the number of defects. Our work instead targets another downside of poor code quality and technical debt, i.e., longer issue resolution times.

Fritz \textit{et al.} developed the degree-of-knowledge model to represent familiarity with source code elements based on a developer's authorship and interaction data~\cite{fritz_degree--knowledge_2014}. The study shows that both constituents are important to correctly model how developers' expertise varies across systems over time. While the authorship information is available in the version control system, the authors developed tools to collect interaction data from the Eclipse development environment.

Baltes and Diehl developed a theory of software development expertise with a focus on programming~\cite{baltes_towards_2018}. Using grounded theory, they developed the theory based on a literature review and a mixed-methods survey with 335 developers. Various factors foster or hinder the formation of expertise, and it may also decline over time. The authors' data shows that experience is not necessarily related to expertise, which motivates us to study both experience and ownership in this work.

Ostrand \textit{et al.} investigated whether files changed by certain developers systematically contained more faults~\cite{ostrand_programmer-based_2010}. They studied this longitudinally for 16 consecutive quarterly releases of a large AT\&T provisioning system implemented in Java. Their ambition was to improve the prediction accuracy of industrial defect prediction models, but adding individual developers' past performance did not add predictive power. On the other hand, they found evidence of persistent variation regarding fault proneness of developers' code contributions. Assuming that developers' system understanding increases over time, one might have guessed that the fault proneness should decrease. Ostrand \textit{et al.} speculate that more seasoned developers might be assigned more difficult implementation tasks. Another possible explanation is that the system under development grows increasingly complex over time.

\section{Method} \label{sec:method}
This study follows the draft version of the empirical standard for repository mining\footnote{\url{https://github.com/acmsigsoft/EmpiricalStandards/blob/master/docs/RepositoryMining.md} (2022-05-09, Latest commit dbe3ad1)}. Repository mining is the appropriate standard as we use automated techniques to extract data from source code repositories and issue management systems followed by quantitative analysis. As an observational study, our focus is on discovering meaningful patterns~\cite{ayala_use_2022}. %We sampled 40 repositories among CodeScene's customers -- different from our previous study~\cite{tornhill_code_2022}.

\subsection{Research Questions and Variables} \label{sec:rqs}
Our goal is to evaluate how developers' experience and ownership are related to issue resolution time in proprietary software development. We are particularly interested in the differences between high- and low-quality source code. To investigate this, we consider time-in-development as the dependent variable and the following three independent variables:

\begin{description}
    \item[Project experience] How long a developer has been involved in the development of the system. Experience increases over time and we measure it as the time since a developer made the first commit to the repository. To allow cross-project analysis, we normalize the measure across all repositories.
    \item[Code ownership] The fraction of touches a developer has made to a source code file. Ownership can fluctuate over time. In parts of this study, we treat ownership as an ordinal variable with four levels: 
    \begin{itemize}
    \item Marginal ownership [$<0.1$]
    \item Minor ownership [0.1--0.49]
    \item Major ownership [0.5--0.9]
    \item Dominant ownership [$>0.9$]    
    \end{itemize}
    When a developer has a marginal/minor/major/dominant ownership of a file, we refer to the developer as a marginal/minor/major/dominant owner.
    \item[Code Health] Measures the code quality of an individual file. We treat it as an ordinal variable using the categories provided by CodeScene, i.e., healthy, warning, and alert.
\end{description} 

Based on our goal, we set out to explore three Research Questions (RQ) in the context of proprietary software development projects: 

\begin{itemize}
\item[RQ1] How is ownership distributed for evolving software?
\item[RQ2] How does a developer's project experience correlate with issue resolution times?
\item[RQ3] How do issue resolution times vary with code quality for changes made by marginal owners?
\end{itemize}

The rationales behind our RQs are as follows. RQ1 explores if collective code ownership, as recommended in agile development, is common in proprietary development. Measuring ownership is also a prerequisite for RQ3. RQ2 is motivated by the assumption that seasoned developers resolve issues faster. In this work, we explore whether a naive approach to measuring this supports this assumption. RQ3 originates in our assumption that marginal owners struggle more with code comprehension in the presence of technical debt. We hypothesize that it takes a longer time for a developer less acquainted with a source code file to resolve issues if the code itself is of low quality. Moreover, we expect to find a bigger difference for corrective maintenance, i.e., bug fixes, than software evolution, i.e., adding new features. The reason is that bug fixes represent unplanned work whereas feature implementation is planned -- we believe that the differences in issue resolution times would be amplified by poor code quality when marginal owners need to make unprepared changes.

Inevitably, many confounding factors are at play in software engineering. For this study, we believe that the complexity of individual issues is the most important one to control. As we cannot measure the complexity directly, we use the size of the change that resolved the issue as a proxy. This is in line with previous work~\cite{atkins_measuring_2000,kononenko_code_2016,motwani_automated_2018}. Based on the change size measured in LoC, we use blocking to mitigate the influence of complexity on the issue resolution time. We use the following three blocks: Small ($<5$ LoC), Medium (6--19 LoC), and Large ($\geq 20$ LoC).

\subsection{Data Collection and Preprocessing} \label{sec:data}
We selected 40 proprietary repositories from 18 companies for inclusion in this study.
The sample contains data from a mix of CodeScene customers, representing a varied set of proprietary software engineering projects.
For confidentiality reasons, we cannot disclose the exact selection criteria -- but we stress that it is a different set compared to the Code Red paper~\cite{tornhill_code_2022}.
Productivity data and defect numbers are sensitive, but we stress that all data owners have provided their consent for us to mine their repositories for the purpose of research studies, given that the companies remain anonymous and that the data cannot be tracked back to their specific codebase, product, or business entity. 

\begin{figure}
    \centering
    \includegraphics[width=0.49\textwidth]{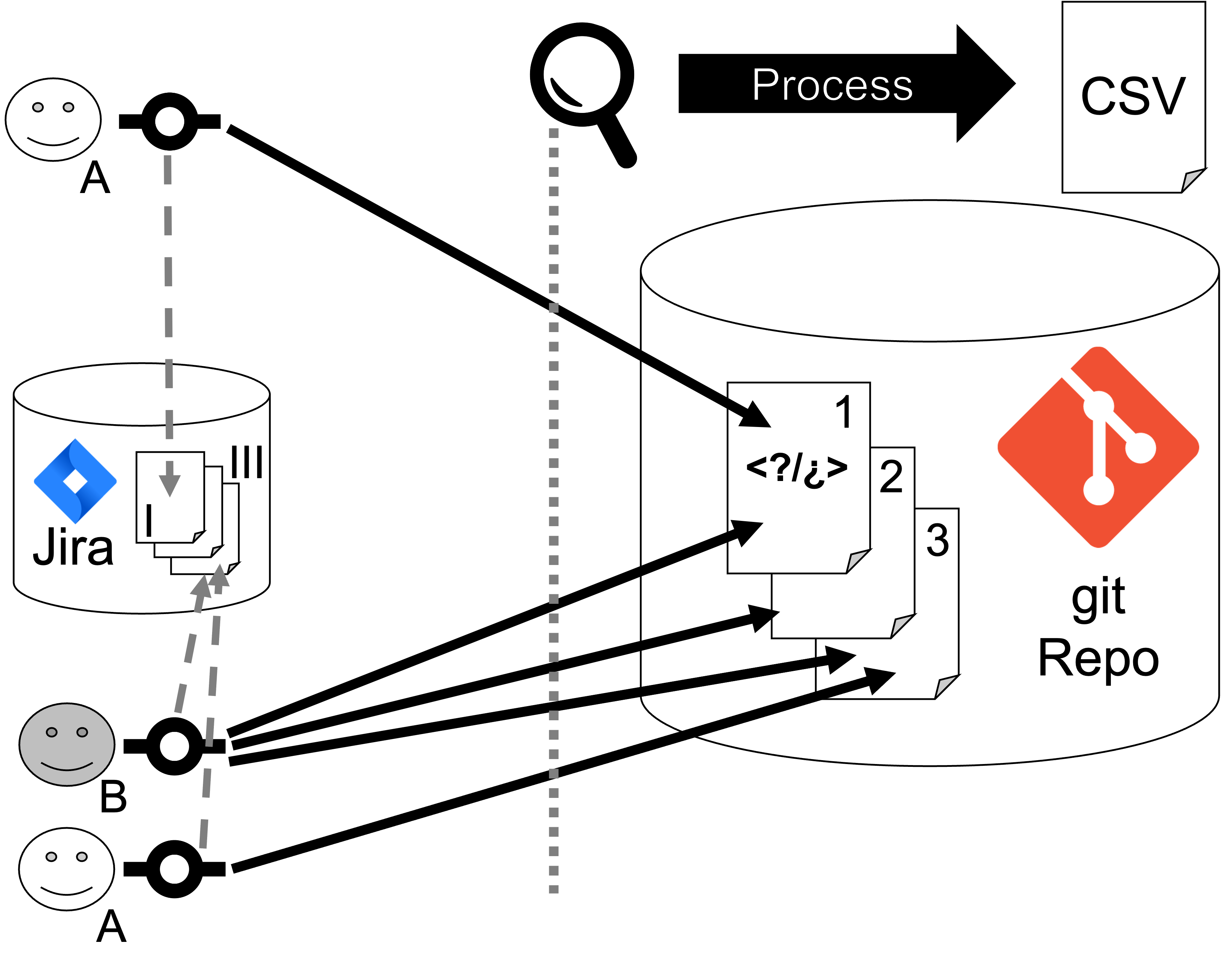}
    \caption{Overview of the data collection.}
    \label{fig:data_coll}
\end{figure}

Fig.~\ref{fig:data_coll} shows an overview of the data collection and the resulting csv-file that is publicly available in the replication package on GitHub~\cite{borg_u_2023}.
All data originate in git repositories and Jira issue trackers.
The magnifying glass and the dashed vertical line illustrates what constitutes rows in the csv-file, i.e., touches~\cite{foucault_usefulness_2015}.
The example in Fig.~\ref{fig:data_coll} shows two developers making three commits, resulting in five touches as follows (starting from the top):

\begin{enumerate}
    \item Dev A's touch of \textbf{File 1} connected to \textbf{Issue I}.
    \item Dev B's touch of \textbf{File 1} connected to \textbf{Issue III}.
    \item Dev B's touch of \textbf{File 2} connected to \textbf{Issue III}.
    \item Dev B's touch of \textbf{File 3} connected to \textbf{Issue III}.
    \item Dev A's touch of \textbf{File 3} connected to \textbf{Issue III}.
\end{enumerate}

The black horizontal arrow in Fig.~\ref{fig:data_coll} represents data processing done to augment the csv-file.
First, the Code Health of individual source code files at commit time is added.
Second, we calculate the developer's code ownership (for the specific file) at commit time.
Third, we calculate the developer's project experience at commit time, scaled to the interval [0-1] across all repositories.

Table~\ref{tab:desc_stats} shows an overview of the 40 development projects analyzed in this study. The table presents the average, standard deviation, lower quartile, upper quartile, and maximum value for the following six metrics:

\begin{itemize}
    \item Size of the codebase measured in thousands of LoC.
    \item Number of files in the git repository.
    \item Number of developers who committed changes.
    \item Number of commits.
    \item Number of Jira issues.
    \item Number of commits connected to a Jira issue.
\end{itemize}

In total, the 40 studied software projects contain 139,869 files with 31,250,579 lines (not restricted to source code). Moreover, the total number of Jira issues is 29,279 and 1,414 developers have made 241,542 commits. Compared to the Code Red paper~\cite{tornhill_code_2022}, the average codebase in our dataset is roughly five times larger whereas the average number of Jira issues per file is 20 times lower.

\begin{table}[]
\caption{Summary statistics of the 40 studied repositories.}
\label{tab:desc_stats}
\begin{tabular}{lccccc|}
\cline{2-6}
\multicolumn{1}{l|}{}                   & Average      & Std        & 25\%      & 75\%   & Max     \\ \hline
\multicolumn{1}{|l|}{Files}             & 3,497         & 4,960        & 541     & 4,346  & 26,339 \\
\multicolumn{1}{|l|}{Lines (in thousands)}               & 781         & 1,515        & 35     & 858  & 7,986 \\
\multicolumn{1}{|l|}{Contributors}      & 35         & 44       & 8         & 39     & 197  \\
\multicolumn{1}{|l|}{Commits}           & 6,038       & 9,052     & 1,017      & 6,943   & 37,173  \\
\multicolumn{1}{|l|}{Jira issues}       & 732.0       & 1,801     & 29      & 455   & 9,671  \\
\multicolumn{1}{|l|}{Commits w. issues}       & 2,623       & 5,754     & 153      & 1,949   & 30,366  \\ \hline
\end{tabular}
\end{table}

The raw dataset, containing 108,235 touches between 2019-09-10 and 2022-12-21, is publicly available as a csv-file~\cite{borg_u_2023}.
As software repositories are known to be noisy~\cite{aranda_secret_2009,al-sabbagh_improving_2022}, we proceed with a number of filtering steps.
An overview is presented below, and all details are described in the Jupyter Notebook provided in the replication package.

\begin{enumerate}
    \item Removal of invalid data. We remove touches for which time-in-development or change size is 0. [-7,006 touches]
    \item Removal of outliers. Touches for which time-in-development or change size is not covered by three standard deviations are removed. [-2,590 touches]
    \item Filter out issues connected to $\geq 50$ touches to mitigate bias of very large issue resolutions. [-43,533 touches] 
    \item Filter out issues with time-in-development $\geq 10,140$ minutes to mitigate bias of very complex issues. This corresponds to a ``mythical man month'' in Sweden 2022, as a homage to Fred Brooks who recently passed away~\cite{brooks_mythical_1995}. [-6,462 touches] 
\end{enumerate}

\begin{figure}
    \centering
    \includegraphics[width=0.45\textwidth]{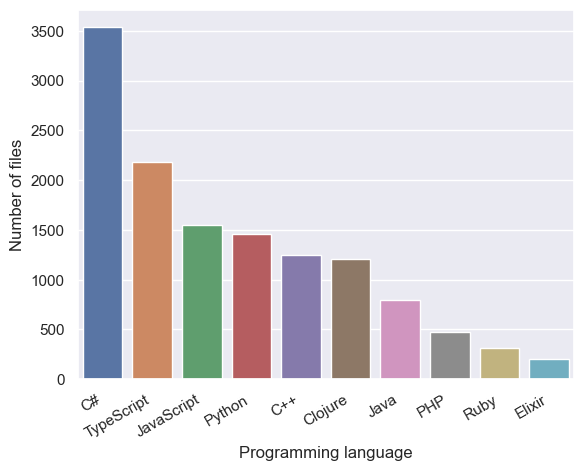}
    \caption{Distribution of the top-10 programming languages.}
    \label{fig:programminglanguages}
\end{figure}

The cleaned dataset used for this study contains 48,644 touches by 457 developers connected to 7,307 Jira issues (2,428 bug reports, 4,879 feature requests).
The touches target 13,433 source code files representing 23 programming languages. Fig.~\ref{fig:programminglanguages} shows the distribution of the most common languages.
We find that C\# is the most common, followed by TypeScript, JavaScript, and Python.
The distributions of the ordinal variables are as follows:

\begin{itemize}
    \item Code Health: 41,361 healthy (84.5\%), 7,100 warning (14.5\%), and 506 alert (1.0\%).
    \item Code Ownership: 14,266 marginal (29.1\%), 14,920 minor (30.5\%), 8,872 major (18.1\%), 10,909 dominant (22.3\%).
    \item Change Size: 21,480 small (44.6\%), 15,699 medium (32.1\%), and 11,428 large (23.3\%).
\end{itemize}

\subsection{Data Analysis}
We explore RQ1 by calculating the relative number of touches developers make to a file. The analysis is only based on descriptive statistics. RQ2 and RQ3, on the other hand, rely on the Python package Pingouin~\cite{vallat_pingouin_2018} and scikit-learn for statistical analysis. 

We investigate correlations in RQ2 by calculating Spearman correlation coefficients ($\rho$). The Spearman correlation is more robust than the Pearson counterpart, which is important due to how skewed our data is toward high values for the normalized project experience. The coefficients' absolute values are interpreted as small between 0.1 and 0.3, medium between 0.3 and 0.5, and large between 0.5 and 1.0. RQ2 is further explored using data visualizations available in the accompanying Jupyter Notebook~\cite{borg_u_2023}.

We study RQ3 using inferential statistics and Kruskal-Wallis H~tests. For readability, we state our null hypotheses related to similar time-in-development for different categories of Code Health in Sec.~\ref{sec:rq3}. When statistical differences are found at $\alpha=0.05$, we conduct Dunn's \textit{post-hoc} test to find out which specific groups' medians are different. Finally, we report effect sizes using Hedges'~g, i.e., the mean difference between two groups divided by the pooled standard deviation. We interpret effect sizes as follows: small between 0.2 and 0.5, medium between 0.5 and 0.8, and large effects from 0.8. %However, when discussing differences in central tendencies, we rely on the more robust median scores rather than the means. 

%To meet the ANOVA pre-conditions, we must ensure the normality of the dependent variable and homoscedasticity between Code Health categories. Fig.~\ref{fig:norm} shows how skewed the data is toward short time-in-development and the distribution after normalization using the Yeo-Johnson algorithm~\cite{yeo_new_2000}. We test the normalized data for homoscedasticity using Levene's test -- details are in the notebook~\cite{borg_u_2023}. 

\begin{figure}
    \centering
    \includegraphics[width=0.35\textwidth]{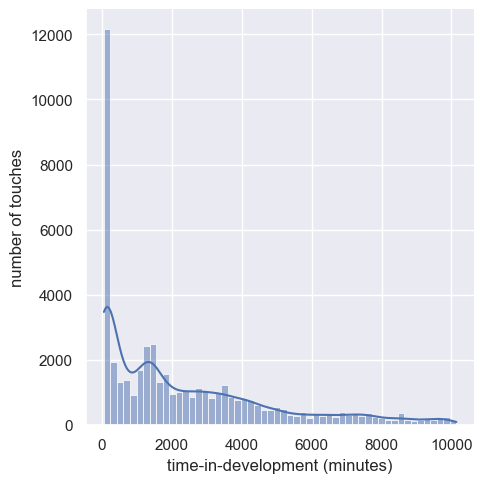}
    \caption{Distribution of time-in-development.}
    \label{fig:norm}
\end{figure}

\section{Results and Discussion} \label{sec:res}
This section reports our results and discusses their implications.

\subsection{RQ1: Ownership Distribution}
\begin{figure}
    \centering
    \includegraphics[width=0.5\textwidth]{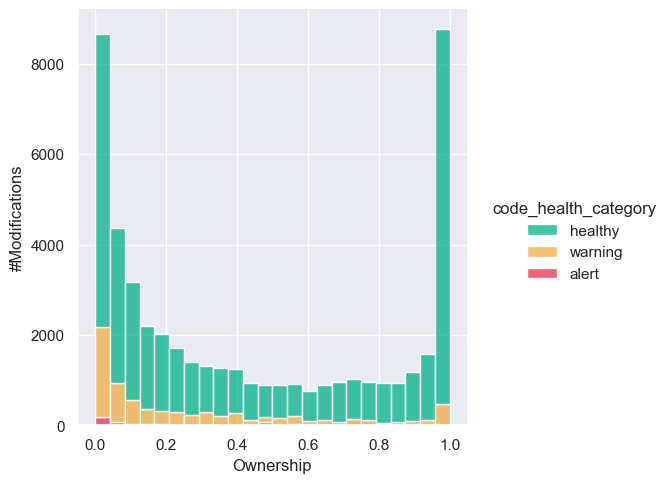}
    \caption{Distribution of code ownership at the time of touch.}
    \label{fig:rq1_res-1}
\end{figure}

Fig.~\ref{fig:rq1_res-1} shows developers' ownership of the target file at change time for 48,644 touches. The data shows a bimodal distribution with modes at the endpoints, i.e., the distribution shows a clear ``U shape'' -- more than 8,000 touches are covered in the first and last bins. Developers' touches frequently target source code files for which they have a marginal ($\le 0.1$) or dominant ($\geq 0.9$) ownership. The distribution is roughly equal for intermediate levels of ownership, but a slight trough can be seen in the center. We capture this observation in an empirically-based rule of thumb:

%\begin{boxB}
\begin{center}
     ``\textit{U owns the code that changes}''
\end{center}
%\end{boxB}

\begin{figure*}
    \centering
    \includegraphics[width=1\textwidth]{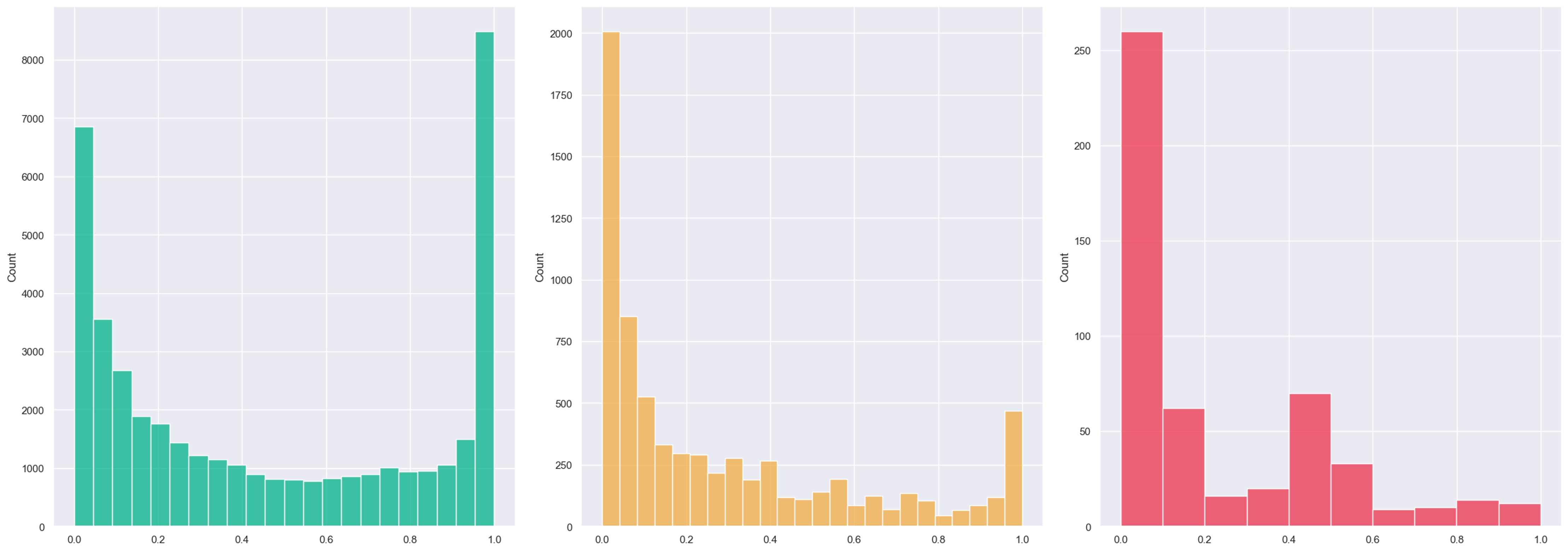}
    \caption{Distribution of code ownership at the time of the touch for healthy, warning, and alert code (from left to right).}
    \label{fig:rq1_res-2}
\end{figure*}

The colors in Figures~\ref{fig:rq1_res-1} and~\ref{fig:rq1_res-2} represent healthy, warning, and alert code quality, respectively. The corresponding distributions of ownership levels vary considerably (cf. Fig.~\ref{fig:rq1_res-2}). For healthy source code, i.e., a majority of the source code analyzed, the ``U shape'' is present. The largest bin is the rightmost, representing a level of ownership $\geq 0.96$. For warning code, the picture is different. The lower end of marginal ownership ($\leq 0.04$) is by far the most common ownership. The pattern for alert code is heavily skewed toward marginal owners and there is no ``U shape'' to be seen.

We find that a majority of touches to files with low-quality source code are made by developers with low levels of ownership. For high-quality source code files, the touches made by marginal owners are balanced by many touches by dominant owners. While we cannot claim any causal relationships, i.e., our data do not show that code quality turns low when there is a lack of touches by dominant owners, there are still reasons for development organizations to monitor such parts of the codebase. Several researchers report that low ownership is related to bug proneness~\cite{bird_dont_2011,greiler_code_2015,jureczko_cross-project_2015,rajapaksha_sqaplanner_2022} -- our findings are in line with these previous studies.

Previous self-assessment surveys show that the adoption of collective code ownership in the software industry varies~\cite{kurapati_agile_2012,rauf_gap_2015,rodriguez_survey_2012}. Many respondents claim to use this practice, but it is clearly not one of the most common practices in the agile toolbox. Our findings corroborate this view, i.e., we find that collective code ownership is not the natural state in proprietary codebases. While we do not know to what extent the companies represented in our dataset strive for this practice, or agile development in general, we can safely say that reaching shared ownership requires a conscious effort. Our focus on touches rather than files also provides a novel perspective on ownership distributions, i.e., we study the most important parts of the codebases -- the files that change. In an IBM case study, Sindghatta \textit{et al.} found that 60\% of the files in a project had multiple owners~\cite{sindhgatta_software_2010}. We show that most changes in 40 repositories happen in files with either a marginal or a dominant owner.\\

%\begin{boxJ}
\textbf{RQ1: U owns the code that changes. Most touches target files with either a marginal or dominant owner. Most touches to files with low-quality source code are made by developers with low levels of ownership.}
%\end{boxJ}

\subsection{RQ2: Experience and Issue Resolution Time}
\begin{figure}
    \centering
    \includegraphics[width=0.45\textwidth]{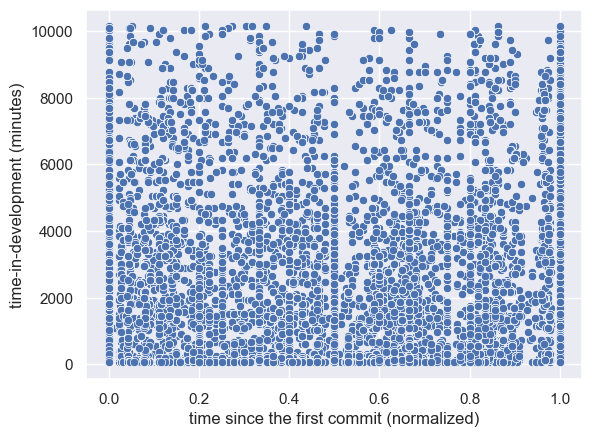}
    \caption{Normalized developer experience vs. time-in-development.}
    \label{fig:rq2_res-corr}
\end{figure}

Fig.~\ref{fig:rq2_res-corr} shows a scatter plot for the relative time since a developer made the first commit to the project vs. the time-in-development for an issue. There is no correlation in the data ($\rho = -0.12$). We also investigated subsets of the data representing different Code Health categories, levels of ownership, and bug fixes only. We found no indication that experience is connected to issue resolution time.

\begin{figure}
    \centering
    \includegraphics[width=0.49\textwidth]{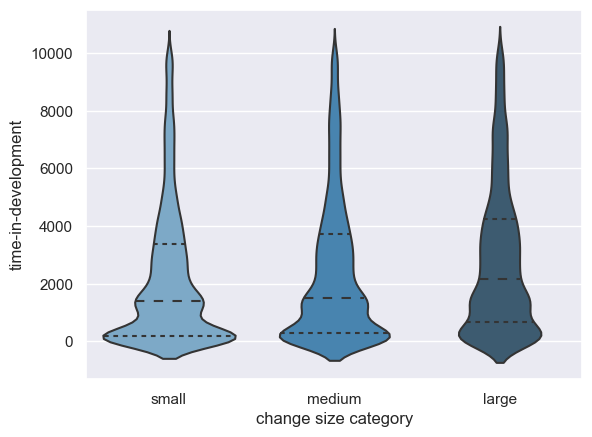}
    \caption{Time-in-development for different touch sizes.}
    \label{fig:rq2_res-violin}
\end{figure}

Do experienced developers work with more challenging issues? This could explain why we do not see any shorter resolution times for more seasoned developers. To explore this assumption, we use change size as a proxy for issue complexity. Fig.~\ref{fig:rq2_res-violin} presents the distribution of time-in-development for small, medium, and large issue resolutions. The data show that many small touches are fast and that the median resolution time for large touches is longer. However, the correlation between touches' number of added LoC and the time-in-development is only $\rho=0.11$. We conclude that the correlation between change size and issue resolution time is very low at best.

\begin{figure}
    \centering
    \includegraphics[width=0.49\textwidth]{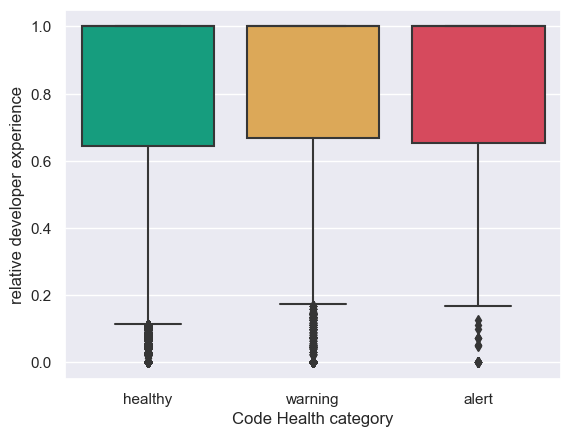}
    \caption{Developers' experience at the time of touch per Code Health category.}
    \label{fig:rq2_res-exp}
\end{figure}

Despite questioning the validity of change size as a complexity proxy, we investigate whether experienced developers tend to make bigger touches. However, we find no correlation ($\rho = -0.04$) in the data. Could it be that experienced developers tend to be assigned to resolve issues in parts of the codebase with challenging technical debt, i.e., warning and alert code? Fig.~\ref{fig:rq2_res-exp} illustrates the relative experience of developers when making changes to source code of different levels of quality. We see no differences and conclude that nothing in our data can explain why developer experience is unrelated to issue resolution time. Software engineering is complex and there are certainly confounding factors at play that we do not capture in our dataset. The conclusion supports previous work that presents experience, without considering the development recency, as a naive measure with limited value~\cite{hattori_refining_2012,baltes_towards_2018}.

%\begin{boxJ}\\
\textbf{RQ2: We find no relationship between developer experience and issue resolution times. Neither change size nor source code quality can explain the absence of the relationship.}
%\end{boxJ}

\subsection{RQ3: Ownership, Issue Resolution Time, and Code Health} \label{sec:rq3}
\begin{figure}
    \centering
    \includegraphics[width=0.49\textwidth]{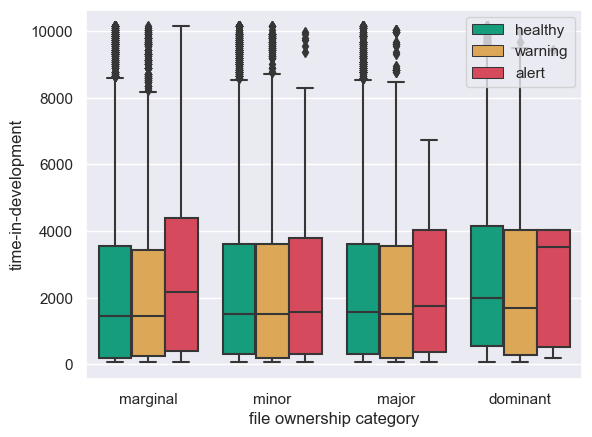}
    \caption{Time-in-development per ownership level for different Code Health categories.}
    \label{fig:rq3-ownership}
\end{figure}

Fig.~\ref{fig:rq3-ownership} shows time-in-development per ownership category for different Code Health categories. We pose four null hypotheses: the time-in-development for different Code Health categories is the same for marginal ($h_{0}1$), minor ($h_{0}2$), major ($h_{0}3$), and dominant ($h_{0}4$) owners. Using Kruskal-Wallis H~tests, we check if there are significant differences between healthy, warning, and alert source code for marginal, minor, major, and dominant owners, respectively. The tests reveal significant differences for marginal owners ($p=0.003$) and dominant owners ($p=0.017$) but not for minor owners ($p=0.536$) and major owners ($p=0.06$). As a result, we reject $h_{0}1$ and $h_{0}4$. In this study, we are particularly interested in the lowest level of ownership and focus our subsequent analysis on changes to files by marginal owners.

\begin{figure}
    \centering
    \includegraphics[width=0.49\textwidth]{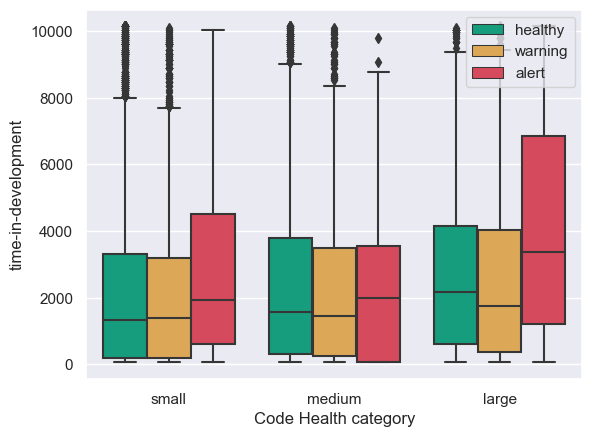}
    \caption{Marginal owners' time-in-development per touch size for different Code Health categories.}
    \label{fig:rq3-size}
\end{figure}

Fig.~\ref{fig:rq3-size} presents the issue resolution time for touches made by marginal owners per touch size for different Code Health categories. Again, we pose null hypotheses accordingly: the time-in-development for touches made by marginal owners for different Code Health categories is the same for small ($h_{0}1S$), medium ($h_{0}1M$), and large ($h_{0}1L$) changes. Kruskal-Wallis H~test lead us to reject $h_{0}1S$ ($p=0.005$) and $h_{0}1L$ ($p=0.016$). We proceed with \textit{post hoc} tests accordingly.

For \textit{small} changes, Dunn's test results in significant differences between healthy and alert code ($p=0.0049$) and warning and alert source code ($p=0.015$), with small effect sizes (Hedge's g is 0.30 and 0.28, respectively). Compared to healthy code, the median time that marginal owners need to make small touches to alert code is 45\% longer. Considering only feature implementation issues, the difference is 96\%. Comparing warning and alert code, the overall difference in median issue resolution time is 39\% (88\% for new feature implementation). 

Looking at \textit{large} touches, Dunn's test gives significant differences between warning and alert source code ($p=0.017$) with a small effect size (Hedge's g is 0.43). Marginal owners need 93\% more time to resolve issues in alert code compared to warning code (98\% for new feature implementation).

Our results provide evidence that resolving issues in low-quality source code requires more time if done by developers with low ownership. The differences are significant for small and large code touches. Our findings support our assumption that developers less familiar with a source code file, such as a developer with low ownership, need additional time to comprehend it. This is in line with a large-scale program comprehension study by Xia \textit{et al.} that investigated developer seniority~\cite{xia_measuring_2018}. Our work, on the other hand, shows similar results for ownership. 

We expected unplanned work, i.e., fixing bugs, to be penalized more than planned work. Our results, however, showed the opposite to be true. The differences when implementing new features is larger than for bug fixes. It is possible that inexperienced developers with low levels of ownership are assigned very simple bugs, perhaps as part of project onboarding activities~\cite{labuschagne_onboarding_2015}. If the bugs are cherry-picked to be particularly easy to solve, then the differences in resolution times could be small despite the overall code quality of the file. On the other hand, the implementation of new features might rarely have such ``mentoring'' Jira issues. Future work should be designed to explain this result.

Our findings add to the growing body of empirical evidence showing the hidden costs of poor code quality~\cite{ampatzoglou_framework_2018,reboucas_de_almeida_tracy_2019,besker_software_2019}. The vast literature on technical debt, which includes tertiary studies~\cite{rios_tertiary_2018,junior_consolidating_2022}, reports that source code debt increases bug proneness and slows down development speed. Moreover, technical debt has a negative impact on developers' job satisfaction~\cite{graziotin_unhappiness_2017} and it is correlated with security vulnerabilities~\cite{siavvas_technical_2022}. In this study, we present another downside: developers with low levels of ownership resolve issues in low-quality source code slower. This has implications for project onboarding, which might require additional time. Furthermore, for organizations struggling with high technical debt, this is particularly troublesome. If the developer churn rate is high because employees quit to instead work with healthier codebases elsewhere, onboarding newcomers must happen more frequently -- and that will require more time.\\

%\begin{boxJ}
\textbf{R3: Marginal owners need a longer time to resolve issues in low-quality source code. Making small and large changes to alert code is 45\% and 93\% slower, respectively. Feature implementation is hampered more than bug fixes.}
%\end{boxJ}

\section{Threats to Validity} \label{sec:threats}
We discuss the threats to validity organized into construct validity, internal validity, conclusion validity, and external validity~\cite{wohlin_experimentation_2012}.

\textbf{Construct validity} refers to the extent to which the studied measures represent what the researchers are considering. Code Health is a proprietary metric that is not entirely transparent. While it has been successfully used in previous research on security vulnerabilities~\cite{al-boghdady_presence_2021} and technical debt~\cite{tornhill_code_2022}, the metric is biased toward a particular perspective on source code quality. Time-in-development, described in our previous work~\cite{tornhill_code_2022}, is used to measure active development time during issue resolution. It is more fine-granular than lead times, but it cannot capture small variations during the development, e.g., context switches, coffee breaks, and recreational web surfing. However, the averaging effects at play in our large dataset support the robustness of our measurements. Finally, we are aware that experience is a questionable proxy for expertise~\cite{baltes_towards_2018}. Thus, we put more emphasis on code ownership in this study. While we identify significant results based on a simple ownership metric, i.e., the fraction of touches, it is possible that more complex metrics from the literature~\cite{greiler_code_2015}, including forgetting effects~\cite{hattori_refining_2012} would have resulted in larger differences. Finally, measuring ownership as the fraction of touches will never tell the whole story -- work practices such as pair programming and discussions between developers cannot be detected in git logs.

\textbf{Internal validity} concerns whether unknown factors might have affected the outcome of the analysis.
As this is an observational repository mining study, its design is inappropriate for drawing causal conclusions~\cite{ayala_use_2022}. Still, we discuss confounding factors that might influence the measured time-in-development. As described in Sec.~\ref{sec:rqs}, we use blocking to mitigate the impact of change size on issue resolution. However, our measure of touch size is simple and considers only modified LoC in the same file. More complex alternatives capturing deltas across files~\cite{atkins_measuring_2000} could be explored in future work. Furthermore, the thresholds we use for blocking into Small, Medium, and Large changes were selected to obtain blocks of roughly equal size. Other thresholds could lead to other results.

Large-scale software engineering is among the most complex collaborative activities humans undertake; thus many other confounders surely exist. To strengthen the validity of our findings, we highlight two factors that could be evaluated in future work. First, code churn~\cite{munson_code_1998,nagappan_use_2005} is often correlated with source code quality and often provides valuable predictive power related to defect proneness. Second, the organizational structure can be related to software quality~\cite{nagappan_influence_2008}. Both these two factors could potentially confound our results. Finally, how we filtered our dataset to roughly 50\% of its original size biases our results. We carefully motivate all filtering steps and, as a sensitivity analysis, we verified that similar patterns are present in the unfiltered dataset. 

\textbf{Conclusion validity} is the degree to which the conclusion we reach is credible and believable. It encompasses both the assumptions we use as part of inferential statistics and whether our interpretations regarding relationships are reasonable. Could it be that we missed relationships in RQ2 or that RQ3 involve fluke findings? Regarding the former, the normalization of developers' project experience across projects resulted in a data skew toward 1.0 -- suggesting truncation effects as can be seen in Fig.~\ref{fig:rq2_res-corr}. The many data points at 1.0 could mean that some repositories were initialized with a large chunk of simultaneous commits. Perhaps a relationship could be found if the data were processed differently.

We found statistically significant differences in RQ3. While there is no registered report available, we embarked on the research with a clear idea to study ownership and code quality. The number of hypotheses was kept small to avoid fishing for p-values lower than the selected confidence level. As a non-parametric test, Kruskal-Wallis does not require neither normality of the dependent variables nor homogeneity of variance. However, the test does assume of random sampling and independence of samples. The independence is violated when mining software repositories for many reasons, e.g., issues are connected~\cite{borg_analyzing_2013} and developers work in teams. %On the other hand, the second assumption is fulfilled through Yeo-Johnson normalization, and we successfully checked the third assumption using Levene's test ($p = 0.23$).

\textbf{External validity} refers to what extent it is possible to generalize the findings. In this study, we are mainly interested in how our conclusions generalize to proprietary software development at large. The 40 repositories analyzed in this study all come from CodeScene customers, which might bias the sample toward organizations that are particularly inclined to value source code quality. While the distributions of code quality in other repositories are different, there is enough alert code in our dataset to detect meaningful patterns. Our sample is obviously very small in relation to the entire proprietary software engineering landscape, but we believe that few academic studies match the number of development contexts represented in our study. Our dataset is cross-organizational, contains both back-end and front-end code implemented in a diverse set of programming languages, and originates in different application domains. Compared to our previous study~\cite{tornhill_code_2022}, dominated by C++, our new dataset leans more toward front-end development given our distribution of languages (cf. Fig.~\ref{fig:programminglanguages}).

While we are focusing our research on proprietary software development, we did analyze whether the rule of thumb ``U owns the code that changes'' generalizes to OSS repositories. Based on an ownership analysis of 301,870 touches extracted from a diverse set of OSS projects, we confirm the presence of the ``U-shape'' -- the endpoints are even more pronounced than in Fig.~\ref{fig:rq1_res-1}. The rule appears largely general, but we do abstain from extrapolating it to best-in-class agile development organizations. We recommend future studies to evaluate whether projects that explicitly claim to apply collective code ownership display another distribution.

\section{Conclusions and Future Work} \label{sec:conc}
Cross-organizational repository mining of proprietary codebases is rare in software engineering research. Our work sheds new light on two phenomena: code ownership and issue resolution. Compared to previous work on ownership, we do not study all repository files but how they change, i.e., the source code that matters the most when pursuing predictable software evolution. 

First, most changes target files with either a marginal or dominant owner. We find a very clear U-shape in the ownership distribution, giving rise to the rule of thumb: ``U owns the code that changes''. While many organizations strive for collective code ownership, the data show low industry adoption of this agile practice. We recommend organizations that actively seek shared ownership to actively measure how it manifests in the codebase. Furthermore, future research could replicate our work and investigate how ownership is distributed in organizations that claim to practice collective code ownership.

Second, we conclude that marginal owners resolve issues slower in low-quality source code. On a similar note, we find that a majority of commits to poor source code are made by marginal owners. The findings have implications for onboarding new developers to codebases with technical debt. While previous research shows how productivity decreases in low-quality code, our work highlights that technical debt impedes marginal owners more. Knowledge distribution through shared code ownership can be a desirable goal, but it will come with a higher cost in low-quality codebases.
Future work should complement our work on code ownership, entirely based on raw touch ratios, with more advanced ownership metrics as well as other approaches to measuring knowledge distribution in software engineering.

\section*{Data Availability}
The anonymized dataset and Jupyter Notebook are available on Zenodo~\cite{borg_u_2023}.

%% file: expertise.bbl

\begin{thebibliography}{54}

%%% the Web address.

\ifx \showCODEN    \undefined \def \showCODEN     #1{\unskip}     \fi
\ifx \showDOI      \undefined \def \showDOI       #1{#1}\fi
\ifx \showISBNx    \undefined \def \showISBNx     #1{\unskip}     \fi
\ifx \showISBNxiii \undefined \def \showISBNxiii  #1{\unskip}     \fi
\ifx \showISSN     \undefined \def \showISSN      #1{\unskip}     \fi
\ifx \showLCCN     \undefined \def \showLCCN      #1{\unskip}     \fi
\ifx \shownote     \undefined \def \shownote      #1{#1}          \fi
\ifx \showarticletitle \undefined \def \showarticletitle #1{#1}   \fi
\ifx \showURL      \undefined \def \showURL       {\relax}        \fi
% invisible to TeX
\providecommand\bibfield[2]{#2}
\providecommand\bibinfo[2]{#2}
\providecommand\natexlab[1]{#1}
\providecommand\showeprint[2][]{arXiv:#2}

\bibitem[\protect\citeauthoryear{Abrantes and Travassos}{Abrantes and
  Travassos}{2011}]%
        {abrantes_common_2011}
\bibfield{author}{\bibinfo{person}{Jose~Fortuna Abrantes} {and}
  \bibinfo{person}{Guilherme~Horta Travassos}.}
  \bibinfo{year}{2011}\natexlab{}.
\newblock \showarticletitle{Common {Agile} {Practices} in {Software}
  {Processes}}. In \bibinfo{booktitle}{\emph{Proc. of the 5th {International}
  {Symposium} on {Empirical} {Software} {Engineering} and {Measurement}}}.
  \bibinfo{pages}{355--358}.
\newblock


\bibitem[\protect\citeauthoryear{Al-Boghdady, Wassif, and El-Ramly}{Al-Boghdady
  et~al\mbox{.}}{2021}]%
        {al-boghdady_presence_2021}
\bibfield{author}{\bibinfo{person}{Abdullah Al-Boghdady},
  \bibinfo{person}{Khaled Wassif}, {and} \bibinfo{person}{Mohammad El-Ramly}.}
  \bibinfo{year}{2021}\natexlab{}.
\newblock \showarticletitle{The {Presence}, {Trends}, and {Causes} of
  {Security} {Vulnerabilities} in {Operating} {Systems} of {IoT}’s
  {Low}-{End} {Devices}}.
\newblock \bibinfo{journal}{\emph{Sensors}} \bibinfo{volume}{21},
  \bibinfo{number}{7} (\bibinfo{year}{2021}), \bibinfo{pages}{2329}.
\newblock


\bibitem[\protect\citeauthoryear{Al-Sabbagh, Staron, and Hebig}{Al-Sabbagh
  et~al\mbox{.}}{2022}]%
        {al-sabbagh_improving_2022}
\bibfield{author}{\bibinfo{person}{Khaled~Walid Al-Sabbagh},
  \bibinfo{person}{Miroslaw Staron}, {and} \bibinfo{person}{Regina Hebig}.}
  \bibinfo{year}{2022}\natexlab{}.
\newblock \showarticletitle{Improving {Test} {Case} {Selection} by {Handling}
  {Class} and {Attribute} {Noise}}.
\newblock \bibinfo{journal}{\emph{Journal of Systems and Software}}
  \bibinfo{volume}{183} (\bibinfo{year}{2022}), \bibinfo{pages}{111093}.
\newblock
\showISSN{0164-1212}


\bibitem[\protect\citeauthoryear{Ampatzoglou, Michailidis, Sarikyriakidis,
  Ampatzoglou, Chatzigeorgiou, and Avgeriou}{Ampatzoglou et~al\mbox{.}}{2018}]%
        {ampatzoglou_framework_2018}
\bibfield{author}{\bibinfo{person}{Areti Ampatzoglou},
  \bibinfo{person}{Alexandros Michailidis}, \bibinfo{person}{Christos
  Sarikyriakidis}, \bibinfo{person}{Apostolos Ampatzoglou},
  \bibinfo{person}{Alexander Chatzigeorgiou}, {and} \bibinfo{person}{Paris
  Avgeriou}.} \bibinfo{year}{2018}\natexlab{}.
\newblock \showarticletitle{A {Framework} for {Managing} {Interest} in
  {Technical} {Debt}: {An} {Industrial} {Validation}}. In
  \bibinfo{booktitle}{\emph{Proc. of the 1st {International} {Conference} on
  {Technical} {Debt}}}. \bibinfo{pages}{115--124}.
\newblock
\showISBNx{978-1-4503-5713-5}


\bibitem[\protect\citeauthoryear{Aranda and Venolia}{Aranda and
  Venolia}{2009}]%
        {aranda_secret_2009}
\bibfield{author}{\bibinfo{person}{Jorge Aranda} {and} \bibinfo{person}{Gina
  Venolia}.} \bibinfo{year}{2009}\natexlab{}.
\newblock \showarticletitle{The {Secret} {Life} of {Bugs}: {Going} {Past} the
  {Errors} and {Omissions} in {Software} {Repositories}}. In
  \bibinfo{booktitle}{\emph{Proc. of the 31st {Int}'l. {Conference} on
  {Software} {Engineering}}}. \bibinfo{pages}{298--308}.
\newblock


\bibitem[\protect\citeauthoryear{Atkins, Mockus, and Siy}{Atkins
  et~al\mbox{.}}{2000}]%
        {atkins_measuring_2000}
\bibfield{author}{\bibinfo{person}{David~L. Atkins}, \bibinfo{person}{Audris
  Mockus}, {and} \bibinfo{person}{Harvey~P. Siy}.}
  \bibinfo{year}{2000}\natexlab{}.
\newblock \showarticletitle{Measuring {Technology} {Effects} on {Software}
  {Change} {Cost}}.
\newblock \bibinfo{journal}{\emph{Bell Labs Technical Journal}}
  \bibinfo{volume}{5}, \bibinfo{number}{2} (\bibinfo{year}{2000}),
  \bibinfo{pages}{7--18}.
\newblock
\showISSN{1538-7305}


\bibitem[\protect\citeauthoryear{Ayala, Turhan, Franch, and Juristo}{Ayala
  et~al\mbox{.}}{2022}]%
        {ayala_use_2022}
\bibfield{author}{\bibinfo{person}{Claudia Ayala}, \bibinfo{person}{Burak
  Turhan}, \bibinfo{person}{Xavier Franch}, {and} \bibinfo{person}{Natalia
  Juristo}.} \bibinfo{year}{2022}\natexlab{}.
\newblock \showarticletitle{Use and {Misuse} of the {Term} “{Experiment}”
  in {Mining} {Software} {Repositories} {Research}}.
\newblock \bibinfo{journal}{\emph{IEEE Transactions on Software Engineering}}
  \bibinfo{volume}{48}, \bibinfo{number}{11} (\bibinfo{year}{2022}),
  \bibinfo{pages}{4229--4248}.
\newblock
\showISSN{1939-3520}


\bibitem[\protect\citeauthoryear{Baggen, Correia, Schill, and Visser}{Baggen
  et~al\mbox{.}}{2012}]%
        {baggen_standardized_2012}
\bibfield{author}{\bibinfo{person}{Robert Baggen}, \bibinfo{person}{José~Pedro
  Correia}, \bibinfo{person}{Katrin Schill}, {and} \bibinfo{person}{Joost
  Visser}.} \bibinfo{year}{2012}\natexlab{}.
\newblock \showarticletitle{Standardized {Code} {Quality} {Benchmarking} for
  {Improving} {Software} {Maintainability}}.
\newblock \bibinfo{journal}{\emph{Software Quality Journal}}
  \bibinfo{number}{2} (\bibinfo{year}{2012}), \bibinfo{pages}{287--307}.
\newblock


\bibitem[\protect\citeauthoryear{Bailey, Godbole, Knutson, and Krein}{Bailey
  et~al\mbox{.}}{2013}]%
        {bailey_decade_2013}
\bibfield{author}{\bibinfo{person}{Sabrina~E. Bailey},
  \bibinfo{person}{Sneha~S. Godbole}, \bibinfo{person}{Charles~D. Knutson},
  {and} \bibinfo{person}{Jonathan~L. Krein}.} \bibinfo{year}{2013}\natexlab{}.
\newblock \showarticletitle{A {Decade} of {Conway}'s {Law}: {A} {Literature}
  {Review} from 2003-2012}. In \bibinfo{booktitle}{\emph{Proc. of the 3rd
  {International} {Workshop} on {Replication} in {Empirical} {Software}
  {Engineering} {Research}}}. \bibinfo{pages}{1--14}.
\newblock


\bibitem[\protect\citeauthoryear{Baltes and Diehl}{Baltes and Diehl}{2018}]%
        {baltes_towards_2018}
\bibfield{author}{\bibinfo{person}{Sebastian Baltes} {and}
  \bibinfo{person}{Stephan Diehl}.} \bibinfo{year}{2018}\natexlab{}.
\newblock \showarticletitle{Towards a {Theory} of {Software} {Development}
  {Expertise}}. In \bibinfo{booktitle}{\emph{Proc. of the 26th {ACM} {Joint}
  {Meeting} on {European} {Software} {Engineering} {Conference} and the
  {Symposium} on the {Foundations} of {Software} {Engineering}}}.
  \bibinfo{pages}{187--200}.
\newblock
\showISBNx{978-1-4503-5573-5}


\bibitem[\protect\citeauthoryear{Besker, Martini, and Bosch}{Besker
  et~al\mbox{.}}{2019}]%
        {besker_software_2019}
\bibfield{author}{\bibinfo{person}{Terese Besker}, \bibinfo{person}{Antonio
  Martini}, {and} \bibinfo{person}{Jan Bosch}.}
  \bibinfo{year}{2019}\natexlab{}.
\newblock \showarticletitle{Software {Developer} {Productivity} {Loss} {Due} to
  {Technical} {Debt}: {A} {Replication} and {Extension} {Study} {Examining}
  {Developers}’ {Development} {Work}}.
\newblock \bibinfo{journal}{\emph{Journal of Systems and Software}}
  \bibinfo{volume}{156} (\bibinfo{year}{2019}), \bibinfo{pages}{41--61}.
\newblock


\bibitem[\protect\citeauthoryear{Bird, Nagappan, Murphy, Gall, and
  Devanbu}{Bird et~al\mbox{.}}{2011}]%
        {bird_dont_2011}
\bibfield{author}{\bibinfo{person}{Christian Bird}, \bibinfo{person}{Nachi
  Nagappan}, \bibinfo{person}{Brendan Murphy}, \bibinfo{person}{Harald Gall},
  {and} \bibinfo{person}{Premkumar Devanbu}.} \bibinfo{year}{2011}\natexlab{}.
\newblock \showarticletitle{Don’t {Touch} {My} {Code}! {Examining} the
  {Effects} of {Ownership} on {Software} {Quality}}. In
  \bibinfo{booktitle}{\emph{Proc. of the the 8th {Joint} {Meeting} of the
  {European} {Software} {Engineering} {Conference} and the {Symposium} on {The}
  {Foundations} of {Software} {Engineering}}}.
\newblock


\bibitem[\protect\citeauthoryear{Boehm}{Boehm}{1983}]%
        {boehm_seven_1983}
\bibfield{author}{\bibinfo{person}{Barry~W. Boehm}.}
  \bibinfo{year}{1983}\natexlab{}.
\newblock \showarticletitle{Seven basic principles of {Software}
  {Engineering}}.
\newblock \bibinfo{journal}{\emph{Journal of Systems and Software}}
  \bibinfo{volume}{3}, \bibinfo{number}{1} (\bibinfo{year}{1983}),
  \bibinfo{pages}{3--24}.
\newblock
\showISSN{0164-1212}


\bibitem[\protect\citeauthoryear{Borg, Pfahl, and Runeson}{Borg
  et~al\mbox{.}}{2013}]%
        {borg_analyzing_2013}
\bibfield{author}{\bibinfo{person}{Markus Borg}, \bibinfo{person}{Dietmar
  Pfahl}, {and} \bibinfo{person}{Per Runeson}.}
  \bibinfo{year}{2013}\natexlab{}.
\newblock \showarticletitle{Analyzing {Networks} of {Issue} {Reports}}. In
  \bibinfo{booktitle}{\emph{Proc. of the 17th {European} {Conference} on
  {Software} {Maintenance} and {Reengineering}}}. \bibinfo{pages}{79--88}.
\newblock


\bibitem[\protect\citeauthoryear{Borg, Tornhill, and Mones}{Borg
  et~al\mbox{.}}{2023}]%
        {borg_u_2023}
\bibfield{author}{\bibinfo{person}{Markus Borg}, \bibinfo{person}{Adam
  Tornhill}, {and} \bibinfo{person}{Enys Mones}.}
  \bibinfo{year}{2023}\natexlab{}.
\newblock \bibinfo{title}{U {Owns} the {Code}, {Replication} {Package}}.
\newblock
\newblock
\urldef\tempurl%
\url{https://doi.org/10.5281/zenodo.7852985}
\showURL{%
\tempurl}


\bibitem[\protect\citeauthoryear{Brooks}{Brooks}{1995}]%
        {brooks_mythical_1995}
\bibfield{author}{\bibinfo{person}{Frederick Brooks}.}
  \bibinfo{year}{1995}\natexlab{}.
\newblock \bibinfo{booktitle}{\emph{The {Mythical} {Man}-{Month}: {Essays} {On}
  {Software} {Engineering}}}.
\newblock \bibinfo{publisher}{Pearson Education}.
\newblock
\showISBNx{978-0-13-211916-0}


\bibitem[\protect\citeauthoryear{Fenton}{Fenton}{1994}]%
        {fenton_software_1994}
\bibfield{author}{\bibinfo{person}{Norman Fenton}.}
  \bibinfo{year}{1994}\natexlab{}.
\newblock \showarticletitle{Software {Measurement}: {A} {Necessary}
  {Scientific} {Basis}}.
\newblock \bibinfo{journal}{\emph{IEEE Transactions on Software Engineering}}
  \bibinfo{volume}{20}, \bibinfo{number}{3} (\bibinfo{year}{1994}),
  \bibinfo{pages}{199--206}.
\newblock


\bibitem[\protect\citeauthoryear{Foucault, Falleri, and Blanc}{Foucault
  et~al\mbox{.}}{2014}]%
        {foucault_code_2014}
\bibfield{author}{\bibinfo{person}{Matthieu Foucault},
  \bibinfo{person}{Jean-Rémy Falleri}, {and} \bibinfo{person}{Xavier Blanc}.}
  \bibinfo{year}{2014}\natexlab{}.
\newblock \showarticletitle{Code {Ownership} in {Open}-{Source} {Software}}. In
  \bibinfo{booktitle}{\emph{Proc. of the 18th {International} {Conference} on
  {Evaluation} and {Assessment} in {Software} {Engineering}}}.
  \bibinfo{pages}{1--9}.
\newblock
\showISBNx{978-1-4503-2476-2}


\bibitem[\protect\citeauthoryear{Foucault, Teyton, Lo, Blanc, and
  Falleri}{Foucault et~al\mbox{.}}{2015}]%
        {foucault_usefulness_2015}
\bibfield{author}{\bibinfo{person}{Matthieu Foucault}, \bibinfo{person}{Cedric
  Teyton}, \bibinfo{person}{David Lo}, \bibinfo{person}{Xavier Blanc}, {and}
  \bibinfo{person}{Jean-Remy Falleri}.} \bibinfo{year}{2015}\natexlab{}.
\newblock \showarticletitle{On the {Usefulness} of {Ownership} {Metrics} in
  {Open}-{Source} {Software} {Projects}}.
\newblock \bibinfo{journal}{\emph{Information and Software Technology}}
  \bibinfo{volume}{64} (\bibinfo{year}{2015}), \bibinfo{pages}{102--112}.
\newblock
\showISSN{0950-5849}


\bibitem[\protect\citeauthoryear{Fritz, Murphy, Murphy-Hill, Ou, and
  Hill}{Fritz et~al\mbox{.}}{2014}]%
        {fritz_degree--knowledge_2014}
\bibfield{author}{\bibinfo{person}{Thomas Fritz}, \bibinfo{person}{Gail~C.
  Murphy}, \bibinfo{person}{Emerson Murphy-Hill}, \bibinfo{person}{Jingwen Ou},
  {and} \bibinfo{person}{Emily Hill}.} \bibinfo{year}{2014}\natexlab{}.
\newblock \showarticletitle{Degree-of-{Knowledge}: {Modeling} a {Developer}'s
  {Knowledge} of {Code}}.
\newblock \bibinfo{journal}{\emph{ACM Transactions on Software Engineering and
  Methodology}} \bibinfo{volume}{23}, \bibinfo{number}{2}
  (\bibinfo{year}{2014}), \bibinfo{pages}{14:1--14:42}.
\newblock
\showISSN{1049-331X}


\bibitem[\protect\citeauthoryear{Graziotin, Fagerholm, Wang, and
  Abrahamsson}{Graziotin et~al\mbox{.}}{2017}]%
        {graziotin_unhappiness_2017}
\bibfield{author}{\bibinfo{person}{Daniel Graziotin}, \bibinfo{person}{Fabian
  Fagerholm}, \bibinfo{person}{Xiaofeng Wang}, {and} \bibinfo{person}{Pekka
  Abrahamsson}.} \bibinfo{year}{2017}\natexlab{}.
\newblock \showarticletitle{On the {Unhappiness} of {Software} {Developers}}.
  In \bibinfo{booktitle}{\emph{Proc. of the 21st {International} {Conference}
  on {Evaluation} and {Assessment} in {Software} {Engineering}}}.
  \bibinfo{pages}{324--333}.
\newblock
\showISBNx{978-1-4503-4804-1}


\bibitem[\protect\citeauthoryear{Greiler, Herzig, and Czerwonka}{Greiler
  et~al\mbox{.}}{2015}]%
        {greiler_code_2015}
\bibfield{author}{\bibinfo{person}{Michaela Greiler}, \bibinfo{person}{Kim
  Herzig}, {and} \bibinfo{person}{Jacek Czerwonka}.}
  \bibinfo{year}{2015}\natexlab{}.
\newblock \showarticletitle{Code {Ownership} and {Software} {Quality}: {A}
  {Replication} {Study}}. In \bibinfo{booktitle}{\emph{Proc. of the 12th
  {Working} {Conference} on {Mining} {Software} {Repositories}}}.
  \bibinfo{pages}{2--12}.
\newblock


\bibitem[\protect\citeauthoryear{Hattori, Lanza, and Robbes}{Hattori
  et~al\mbox{.}}{2012}]%
        {hattori_refining_2012}
\bibfield{author}{\bibinfo{person}{Lile~Palma Hattori},
  \bibinfo{person}{Michele Lanza}, {and} \bibinfo{person}{Romain Robbes}.}
  \bibinfo{year}{2012}\natexlab{}.
\newblock \showarticletitle{Refining {Code} {Ownership} {With} {Synchronous}
  {Changes}}.
\newblock \bibinfo{journal}{\emph{Empirical Software Engineering}}
  \bibinfo{volume}{17}, \bibinfo{number}{4} (\bibinfo{year}{2012}),
  \bibinfo{pages}{467--499}.
\newblock
\showISSN{1573-7616}


\bibitem[\protect\citeauthoryear{{International Organization for
  Standardization}}{{International Organization for Standardization}}{2021}]%
        {international_organization_for_standardization_information_2021}
\bibfield{author}{\bibinfo{person}{{International Organization for
  Standardization}}.} \bibinfo{year}{2021}\natexlab{}.
\newblock \bibinfo{booktitle}{\emph{Information {Technology} - {Software}
  {Measurement} - {Software} {Quality} {Measurement} - {Automated} {Source}
  {Code} {Quality} {Measures}}}.
\newblock \bibinfo{type}{{T}echnical {R}eport} ISO/IEC 5055:2021.
  \bibinfo{institution}{International Organization for Standardization}.
\newblock


\bibitem[\protect\citeauthoryear{Jorgensen and Shepperd}{Jorgensen and
  Shepperd}{2007}]%
        {jorgensen_systematic_2007}
\bibfield{author}{\bibinfo{person}{Magne Jorgensen} {and}
  \bibinfo{person}{Martin Shepperd}.} \bibinfo{year}{2007}\natexlab{}.
\newblock \showarticletitle{A {Systematic} {Review} of {Software} {Development}
  {Cost} {Estimation} {Studies}}.
\newblock \bibinfo{journal}{\emph{IEEE Transactions on Software Engineering}}
  \bibinfo{volume}{33}, \bibinfo{number}{1} (\bibinfo{year}{2007}),
  \bibinfo{pages}{33--53}.
\newblock
\showISSN{1939-3520}


\bibitem[\protect\citeauthoryear{Junior and Travassos}{Junior and
  Travassos}{2022}]%
        {junior_consolidating_2022}
\bibfield{author}{\bibinfo{person}{Helvio~Jeronimo Junior} {and}
  \bibinfo{person}{Guilherme~Horta Travassos}.}
  \bibinfo{year}{2022}\natexlab{}.
\newblock \showarticletitle{Consolidating a {Common} {Perspective} on
  {Technical} {Debt} and its {Management} {Through} a {Tertiary} {Study}}.
\newblock \bibinfo{journal}{\emph{Information and Software Technology}}
  \bibinfo{volume}{149} (\bibinfo{year}{2022}), \bibinfo{pages}{106964}.
\newblock
\showISSN{0950-5849}


\bibitem[\protect\citeauthoryear{Jureczko and Madeyski}{Jureczko and
  Madeyski}{2015}]%
        {jureczko_cross-project_2015}
\bibfield{author}{\bibinfo{person}{Marian Jureczko} {and} \bibinfo{person}{Lech
  Madeyski}.} \bibinfo{year}{2015}\natexlab{}.
\newblock \showarticletitle{Cross-{Project} {Defect} {Prediction} with
  {Respect} to {Code} {Ownership} {Model}: {An} {Empirical} {Study}}.
\newblock \bibinfo{journal}{\emph{e-Informatica Software Engineering Journal}}
  \bibinfo{volume}{9}, \bibinfo{number}{1} (\bibinfo{year}{2015}).
\newblock


\bibitem[\protect\citeauthoryear{Kononenko, Baysal, and Godfrey}{Kononenko
  et~al\mbox{.}}{2016}]%
        {kononenko_code_2016}
\bibfield{author}{\bibinfo{person}{Oleksii Kononenko}, \bibinfo{person}{Olga
  Baysal}, {and} \bibinfo{person}{Michael~W. Godfrey}.}
  \bibinfo{year}{2016}\natexlab{}.
\newblock \showarticletitle{Code {Review} {Quality}: {How} {Developers} {See}
  {It}}. In \bibinfo{booktitle}{\emph{Proc. of the 38th {International}
  {Conference} on {Software} {Engineering}}}. \bibinfo{pages}{1028--1038}.
\newblock


\bibitem[\protect\citeauthoryear{Kurapati, Manyam, and Petersen}{Kurapati
  et~al\mbox{.}}{2012}]%
        {kurapati_agile_2012}
\bibfield{author}{\bibinfo{person}{Narendra Kurapati}, \bibinfo{person}{Venkata
  Sarath~Chandra Manyam}, {and} \bibinfo{person}{Kai Petersen}.}
  \bibinfo{year}{2012}\natexlab{}.
\newblock \showarticletitle{Agile {Software} {Development} {Practice}
  {Adoption} {Survey}}. In \bibinfo{booktitle}{\emph{Agile {Processes} in
  {Software} {Engineering} and {Extreme} {Programming}}}
  \emph{(\bibinfo{series}{Lecture {Notes} in {Business} {Information}
  {Processing}})}, \bibfield{editor}{\bibinfo{person}{Claes Wohlin}} (Ed.).
  \bibinfo{publisher}{Springer}, \bibinfo{address}{Berlin, Heidelberg},
  \bibinfo{pages}{16--30}.
\newblock
\showISBNx{978-3-642-30350-0}


\bibitem[\protect\citeauthoryear{Labuschagne and Holmes}{Labuschagne and
  Holmes}{2015}]%
        {labuschagne_onboarding_2015}
\bibfield{author}{\bibinfo{person}{Adriaan Labuschagne} {and}
  \bibinfo{person}{Reid Holmes}.} \bibinfo{year}{2015}\natexlab{}.
\newblock \showarticletitle{Do {Onboarding} {Programs} {Work}?}. In
  \bibinfo{booktitle}{\emph{Proc. of the 12th {Working} {Conference} on
  {Mining} {Software} {Repositories}}}. \bibinfo{pages}{381--385}.
\newblock


\bibitem[\protect\citeauthoryear{Lacerda, Petrillo, Pimenta, and
  Gueheneuc}{Lacerda et~al\mbox{.}}{2020}]%
        {lacerda_code_2020}
\bibfield{author}{\bibinfo{person}{Guilherme Lacerda}, \bibinfo{person}{Fabio
  Petrillo}, \bibinfo{person}{Marcelo Pimenta}, {and}
  \bibinfo{person}{Yann~Gael Gueheneuc}.} \bibinfo{year}{2020}\natexlab{}.
\newblock \showarticletitle{Code {Smells} and {Refactoring}: {A} {Tertiary}
  {Systematic} {Review} of {Challenges} and {Observations}}.
\newblock \bibinfo{journal}{\emph{Journal of Systems and Software}}
  \bibinfo{volume}{167} (\bibinfo{year}{2020}), \bibinfo{pages}{110610}.
\newblock


\bibitem[\protect\citeauthoryear{Malgonde and Chari}{Malgonde and
  Chari}{2019}]%
        {malgonde_ensemble-based_2019}
\bibfield{author}{\bibinfo{person}{Onkar Malgonde} {and}
  \bibinfo{person}{Kaushal Chari}.} \bibinfo{year}{2019}\natexlab{}.
\newblock \showarticletitle{An {Ensemble}-{Based} {Model} for {Predicting}
  {Agile} {Software} {Development} {Effort}}.
\newblock \bibinfo{journal}{\emph{Empirical Software Engineering}}
  \bibinfo{volume}{24}, \bibinfo{number}{2} (\bibinfo{year}{2019}),
  \bibinfo{pages}{1017--1055}.
\newblock
\showISSN{1573-7616}


\bibitem[\protect\citeauthoryear{Motwani, Sankaranarayanan, Just, and
  Brun}{Motwani et~al\mbox{.}}{2018}]%
        {motwani_automated_2018}
\bibfield{author}{\bibinfo{person}{Manish Motwani}, \bibinfo{person}{Sandhya
  Sankaranarayanan}, \bibinfo{person}{Rene Just}, {and} \bibinfo{person}{Yuriy
  Brun}.} \bibinfo{year}{2018}\natexlab{}.
\newblock \showarticletitle{Do {Automated} {Program} {Repair} {Techniques}
  {Repair} {Hard} and {Important} {Bugs}?}
\newblock \bibinfo{journal}{\emph{Empirical Software Engineering}}
  \bibinfo{volume}{23}, \bibinfo{number}{5} (\bibinfo{year}{2018}),
  \bibinfo{pages}{2901--2947}.
\newblock
\showISSN{1573-7616}


\bibitem[\protect\citeauthoryear{Munson and Elbaum}{Munson and Elbaum}{1998}]%
        {munson_code_1998}
\bibfield{author}{\bibinfo{person}{J.C. Munson} {and} \bibinfo{person}{S.G.
  Elbaum}.} \bibinfo{year}{1998}\natexlab{}.
\newblock \showarticletitle{Code {Churn}: {A} {Measure} for {Estimating} the
  {Impact} of {Code} {Change}}. In \bibinfo{booktitle}{\emph{Proc. of the
  {International} {Conference} on {Software} {Maintenance}}}.
  \bibinfo{pages}{24--31}.
\newblock


\bibitem[\protect\citeauthoryear{Naedele, Chen, Kazman, Cai, Xiao, and
  Silva}{Naedele et~al\mbox{.}}{2015}]%
        {naedele_manufacturing_2015}
\bibfield{author}{\bibinfo{person}{Martin Naedele}, \bibinfo{person}{Hong-Mei
  Chen}, \bibinfo{person}{Rick Kazman}, \bibinfo{person}{Yuanfang Cai},
  \bibinfo{person}{Lu Xiao}, {and} \bibinfo{person}{Carlos V.~A. Silva}.}
  \bibinfo{year}{2015}\natexlab{}.
\newblock \showarticletitle{Manufacturing {Execution} {Systems}: {A} {Vision}
  for {Managing} {Software} {Development}}.
\newblock \bibinfo{journal}{\emph{Journal of Systems and Software}}
  \bibinfo{volume}{101} (\bibinfo{year}{2015}), \bibinfo{pages}{59--68}.
\newblock
\showISSN{0164-1212}


\bibitem[\protect\citeauthoryear{Nagappan and Ball}{Nagappan and Ball}{2005}]%
        {nagappan_use_2005}
\bibfield{author}{\bibinfo{person}{Nachiappan Nagappan} {and}
  \bibinfo{person}{Thomas Ball}.} \bibinfo{year}{2005}\natexlab{}.
\newblock \showarticletitle{Use of {Relative} {Code} {Churn} {Measures} to
  {Predict} {System} {Defect} {Density}}. In \bibinfo{booktitle}{\emph{Proc. of
  the 27th {International} {Conference} on {Software} {Engineering}}}.
  \bibinfo{pages}{284--292}.
\newblock
\showISBNx{978-1-58113-963-1}


\bibitem[\protect\citeauthoryear{Nagappan, Murphy, and Basili}{Nagappan
  et~al\mbox{.}}{2008}]%
        {nagappan_influence_2008}
\bibfield{author}{\bibinfo{person}{Nachiappan Nagappan},
  \bibinfo{person}{Brendan Murphy}, {and} \bibinfo{person}{Victor Basili}.}
  \bibinfo{year}{2008}\natexlab{}.
\newblock \showarticletitle{The {Influence} of {Organizational} {Structure} on
  {Software} {Quality}}. In \bibinfo{booktitle}{\emph{Proc. of the 30th
  {International} {Conference} on {Software} {Engineering}}}.
  \bibinfo{pages}{521--530}.
\newblock


\bibitem[\protect\citeauthoryear{Nunez-Varela, Perez-Gonzalez, Martinez-Perez,
  and Soubervielle-Montalvo}{Nunez-Varela et~al\mbox{.}}{2017}]%
        {nunez-varela_source_2017}
\bibfield{author}{\bibinfo{person}{Alberto~S. Nunez-Varela},
  \bibinfo{person}{Hector~G. Perez-Gonzalez}, \bibinfo{person}{Francisco~E.
  Martinez-Perez}, {and} \bibinfo{person}{Carlos Soubervielle-Montalvo}.}
  \bibinfo{year}{2017}\natexlab{}.
\newblock \showarticletitle{Source {Code} {Metrics}: {A} {Systematic} {Mapping}
  {Study}}.
\newblock \bibinfo{journal}{\emph{Journal of Systems and Software}}
  \bibinfo{volume}{128} (\bibinfo{year}{2017}), \bibinfo{pages}{164--197}.
\newblock
\showISSN{0164-1212}


\bibitem[\protect\citeauthoryear{Ostrand, Weyuker, and Bell}{Ostrand
  et~al\mbox{.}}{2010}]%
        {ostrand_programmer-based_2010}
\bibfield{author}{\bibinfo{person}{Thomas~J. Ostrand},
  \bibinfo{person}{Elaine~J. Weyuker}, {and} \bibinfo{person}{Robert~M. Bell}.}
  \bibinfo{year}{2010}\natexlab{}.
\newblock \showarticletitle{Programmer-{Based} {Fault} {Prediction}}. In
  \bibinfo{booktitle}{\emph{Proc. of the 6th {International} {Conference} on
  {Predictive} {Models} in {Software} {Engineering}}}. \bibinfo{pages}{1--10}.
\newblock
\showISBNx{978-1-4503-0404-7}


\bibitem[\protect\citeauthoryear{Rahman and Devanbu}{Rahman and
  Devanbu}{2011}]%
        {rahman_ownership_2011}
\bibfield{author}{\bibinfo{person}{Foyzur Rahman} {and}
  \bibinfo{person}{Premkumar Devanbu}.} \bibinfo{year}{2011}\natexlab{}.
\newblock \showarticletitle{Ownership, experience and {Defects}: {A}
  {Fine}-{Grained} {Study} of {Authorship}}. In \bibinfo{booktitle}{\emph{Proc.
  of the 33rd {International} {Conference} on {Software} {Engineering}}}.
  \bibinfo{pages}{491--500}.
\newblock


\bibitem[\protect\citeauthoryear{Rajapaksha, Tantithamthavorn, Jiarpakdee,
  Bergmeir, Grundy, and Buntine}{Rajapaksha et~al\mbox{.}}{2022}]%
        {rajapaksha_sqaplanner_2022}
\bibfield{author}{\bibinfo{person}{Dilini Rajapaksha},
  \bibinfo{person}{Chakkrit Tantithamthavorn}, \bibinfo{person}{Jirayus
  Jiarpakdee}, \bibinfo{person}{Christoph Bergmeir}, \bibinfo{person}{John
  Grundy}, {and} \bibinfo{person}{Wray Buntine}.}
  \bibinfo{year}{2022}\natexlab{}.
\newblock \showarticletitle{{SQAPlanner}: {Generating} {Data}-{Informed}
  {Software} {Quality} {Improvement} {Plans}}.
\newblock \bibinfo{journal}{\emph{IEEE Transactions on Software Engineering}}
  \bibinfo{volume}{48}, \bibinfo{number}{8} (\bibinfo{year}{2022}),
  \bibinfo{pages}{2814--2835}.
\newblock


\bibitem[\protect\citeauthoryear{Rauf and AlGhafees}{Rauf and
  AlGhafees}{2015}]%
        {rauf_gap_2015}
\bibfield{author}{\bibinfo{person}{Abdul Rauf} {and} \bibinfo{person}{Mohammad
  AlGhafees}.} \bibinfo{year}{2015}\natexlab{}.
\newblock \showarticletitle{Gap {Analysis} between {State} of {Practice} and
  {State} of {Art} {Practices} in {Agile} {Software} {Development}}. In
  \bibinfo{booktitle}{\emph{2015 {Agile} {Conference}}}.
  \bibinfo{pages}{102--106}.
\newblock


\bibitem[\protect\citeauthoryear{Reboucas~de Almeida, Treude, and
  Kulesza}{Reboucas~de Almeida et~al\mbox{.}}{2019}]%
        {reboucas_de_almeida_tracy_2019}
\bibfield{author}{\bibinfo{person}{Rodrigo Reboucas~de Almeida},
  \bibinfo{person}{Christoph Treude}, {and} \bibinfo{person}{Uira Kulesza}.}
  \bibinfo{year}{2019}\natexlab{}.
\newblock \showarticletitle{Tracy: {A} {Business}-{Driven} {Technical} {Debt}
  {Prioritization} {Framework}}. In \bibinfo{booktitle}{\emph{Proc. of the 35th
  {International} {Conference} on {Software} {Maintenance} and {Evolution}}}.
  \bibinfo{pages}{181--185}.
\newblock


\bibitem[\protect\citeauthoryear{Riaz, Mendes, and Tempero}{Riaz
  et~al\mbox{.}}{2009}]%
        {riaz_systematic_2009}
\bibfield{author}{\bibinfo{person}{Mehwish Riaz}, \bibinfo{person}{Emilia
  Mendes}, {and} \bibinfo{person}{Ewan Tempero}.}
  \bibinfo{year}{2009}\natexlab{}.
\newblock \showarticletitle{A {Systematic} {Review} of {Software}
  {Maintainability} {Prediction} and {Metrics}}. In
  \bibinfo{booktitle}{\emph{Proc. of the 3rd International {Symposium} on
  {Empirical} {Software} {Engineering} and {Measurement}}}.
  \bibinfo{pages}{367--377}.
\newblock


\bibitem[\protect\citeauthoryear{Rios, Mendonça~Neto, and Spínola}{Rios
  et~al\mbox{.}}{2018}]%
        {rios_tertiary_2018}
\bibfield{author}{\bibinfo{person}{Nicolli Rios}, \bibinfo{person}{Manoel
  Gomes~de Mendonça~Neto}, {and} \bibinfo{person}{Rodrigo~Oliveira Spínola}.}
  \bibinfo{year}{2018}\natexlab{}.
\newblock \showarticletitle{A {Tertiary} {Study} on {Technical} {Debt}:
  {Types}, {Management} {Strategies}, {Research} {Trends}, and {Base}
  {Information} for {Practitioners}}.
\newblock \bibinfo{journal}{\emph{Information and Software Technology}}
  \bibinfo{volume}{102} (\bibinfo{year}{2018}), \bibinfo{pages}{117--145}.
\newblock
\showISSN{0950-5849}


\bibitem[\protect\citeauthoryear{Rodríguez, Markkula, Oivo, and
  Turula}{Rodríguez et~al\mbox{.}}{2012}]%
        {rodriguez_survey_2012}
\bibfield{author}{\bibinfo{person}{Pilar Rodríguez}, \bibinfo{person}{Jouni
  Markkula}, \bibinfo{person}{Markku Oivo}, {and} \bibinfo{person}{Kimmo
  Turula}.} \bibinfo{year}{2012}\natexlab{}.
\newblock \showarticletitle{Survey on {Agile} and {Lean} {Usage} in {Finnish}
  {Software} {Industry}}. In \bibinfo{booktitle}{\emph{Proc. of the 6th
  {International} {Symposium} on {Empirical} {Software} {Engineering} and
  {Measurement}}}. \bibinfo{pages}{139--148}.
\newblock


\bibitem[\protect\citeauthoryear{Sedano, Ralph, and Péraire}{Sedano
  et~al\mbox{.}}{2016}]%
        {sedano_practice_2016}
\bibfield{author}{\bibinfo{person}{Todd Sedano}, \bibinfo{person}{Paul Ralph},
  {and} \bibinfo{person}{Cécile Péraire}.} \bibinfo{year}{2016}\natexlab{}.
\newblock \showarticletitle{Practice and {Perception} of {Team} {Code}
  {Ownership}}. In \bibinfo{booktitle}{\emph{Proc. of the 20th {International}
  {Conference} on {Evaluation} and {Assessment} in {Software} {Engineering}}}.
  \bibinfo{pages}{1--6}.
\newblock
\showISBNx{978-1-4503-3691-8}


\bibitem[\protect\citeauthoryear{Siavvas, Tsoukalas, Jankovic, Kehagias, and
  Tzovaras}{Siavvas et~al\mbox{.}}{2022}]%
        {siavvas_technical_2022}
\bibfield{author}{\bibinfo{person}{Miltiadis Siavvas},
  \bibinfo{person}{Dimitrios Tsoukalas}, \bibinfo{person}{Marija Jankovic},
  \bibinfo{person}{Dionysios Kehagias}, {and} \bibinfo{person}{Dimitrios
  Tzovaras}.} \bibinfo{year}{2022}\natexlab{}.
\newblock \showarticletitle{Technical {Debt} as an {Indicator} of {Software}
  {Security} {Risk}: {A} {Machine} {Learning} {Approach} for {Software}
  {Development} {Enterprises}}.
\newblock \bibinfo{journal}{\emph{Enterprise Information Systems}}
  \bibinfo{volume}{16}, \bibinfo{number}{5} (\bibinfo{year}{2022}),
  \bibinfo{pages}{1824017}.
\newblock
\showISSN{1751-7575}


\bibitem[\protect\citeauthoryear{Sindhgatta, Narendra, and Sengupta}{Sindhgatta
  et~al\mbox{.}}{2010}]%
        {sindhgatta_software_2010}
\bibfield{author}{\bibinfo{person}{Renuka Sindhgatta},
  \bibinfo{person}{Nanjangud~C. Narendra}, {and} \bibinfo{person}{Bikram
  Sengupta}.} \bibinfo{year}{2010}\natexlab{}.
\newblock \showarticletitle{Software {Evolution} in {Agile} {Development}: {A}
  {Case} {Study}}. In \bibinfo{booktitle}{\emph{Proc. of the {International}
  {Conference} {Companion} on {Object} {Oriented} {Programming} {Systems}
  {Languages} and {Applications} {Companion}}}. \bibinfo{pages}{105--114}.
\newblock
\showISBNx{978-1-4503-0240-1}


\bibitem[\protect\citeauthoryear{Tawosi, Moussa, and Sarro}{Tawosi
  et~al\mbox{.}}{2022}]%
        {tawosi_relationship_2022}
\bibfield{author}{\bibinfo{person}{Vali Tawosi}, \bibinfo{person}{Rebecca
  Moussa}, {and} \bibinfo{person}{Federica Sarro}.}
  \bibinfo{year}{2022}\natexlab{}.
\newblock \showarticletitle{On the {Relationship} {Between} {Story} {Points}
  and {Development} {Effort} in {Agile} {Open}-{Source} {Software}}. In
  \bibinfo{booktitle}{\emph{Proc. of the 16th {International} {Symposium} on
  {Empirical} {Software} {Engineering} and {Measurement}}}.
  \bibinfo{pages}{183--194}.
\newblock
\showISBNx{978-1-4503-9427-7}


\bibitem[\protect\citeauthoryear{Tornhill and Borg}{Tornhill and Borg}{2022}]%
        {tornhill_code_2022}
\bibfield{author}{\bibinfo{person}{Adam Tornhill} {and} \bibinfo{person}{Markus
  Borg}.} \bibinfo{year}{2022}\natexlab{}.
\newblock \showarticletitle{Code {Red}: {The} {Business} {Impact} of {Code}
  {Quality} - {A} {Quantitative} {Study} of 39 {Proprietary} {Production}
  {Codebases}}. In \bibinfo{booktitle}{\emph{Proc. of the 5th {International}
  {Conference} on {Technical} {Debt}}}. \bibinfo{pages}{11--20}.
\newblock


\bibitem[\protect\citeauthoryear{Vallat}{Vallat}{2018}]%
        {vallat_pingouin_2018}
\bibfield{author}{\bibinfo{person}{Raphael Vallat}.}
  \bibinfo{year}{2018}\natexlab{}.
\newblock \showarticletitle{Pingouin: {Statistics} in {Python}}.
\newblock \bibinfo{journal}{\emph{Journal of Open Source Software}}
  \bibinfo{volume}{3}, \bibinfo{number}{31} (\bibinfo{year}{2018}),
  \bibinfo{pages}{1026}.
\newblock


\bibitem[\protect\citeauthoryear{Wohlin, Runeson, H\"ost, Ohlsson, Regnell, and
  Wessl\'en}{Wohlin et~al\mbox{.}}{2012}]%
        {wohlin_experimentation_2012}
\bibfield{author}{\bibinfo{person}{C. Wohlin}, \bibinfo{person}{P. Runeson},
  \bibinfo{person}{M. H\"ost}, \bibinfo{person}{M. Ohlsson},
  \bibinfo{person}{B. Regnell}, {and} \bibinfo{person}{A. Wessl\'en}.}
  \bibinfo{year}{2012}\natexlab{}.
\newblock \bibinfo{booktitle}{\emph{Experimentation in {Software}
  {Engineering}: {A} {Practical} {Guide}}}.
\newblock \bibinfo{publisher}{Springer}.
\newblock


\bibitem[\protect\citeauthoryear{Xia, Bao, Lo, Xing, Hassan, and Li}{Xia
  et~al\mbox{.}}{2018}]%
        {xia_measuring_2018}
\bibfield{author}{\bibinfo{person}{Xin Xia}, \bibinfo{person}{Lingfeng Bao},
  \bibinfo{person}{David Lo}, \bibinfo{person}{Zhenchang Xing},
  \bibinfo{person}{Ahmed~E. Hassan}, {and} \bibinfo{person}{Shanping Li}.}
  \bibinfo{year}{2018}\natexlab{}.
\newblock \showarticletitle{Measuring {Program} {Comprehension}: {A}
  {Large}-{Scale} {Field} {Study} with {Professionals}}.
\newblock \bibinfo{journal}{\emph{IEEE Transactions on Software Engineering}}
  \bibinfo{volume}{44}, \bibinfo{number}{10} (\bibinfo{year}{2018}),
  \bibinfo{pages}{951--976}.
\newblock


\end{thebibliography}
